\documentclass[11pt,a4paper]{article}
\usepackage{jheppub}
\usepackage{amsmath}

\title{Penetrating component in cosmic rays.}
\author{Sergey Shaulov}
\affiliation{P.N.Lebedev Physical Institute of Russian Academy of Sciencies,\\
Leninsky prospect 53, Moscow, Russia}
\emailAdd{shaulovsb@lebedev.ru}

\abstract{We present a study of the high energy spectra of hadrons in cores of extensive air showers. These data were obtained for the first time in the hybrid {\it HADRON} experiment (Tien-Shan) by means of a large X-ray emulsion chamber  combined with the shower array. In the local energy interval 3--100 PeV  an increase in the energy of hadrons was found, which means the appearance of a penetrating component.  This component  in our experiment  was observed in the atmosphere  that indicates the presence of a penetrating strongly interacting  component in primary cosmic rays. Along with that, it is worth emphasising that the region where this component is observed coincides with the region of the so-called  knee in the spectrum of cosmic rays. On this basis, a new hypothesis of knee formation can be put forward.}


\begin{document}
\maketitle

\section{Introduction.}

The first indication  of a penetrating component in cosmic rays were obtained in the experiment with the calorimeter \cite{dremin}. In 1980, it was found that at the hadron energies of 100 TeV, the absorption length of cascade in the lead calorimeter increases from 500 to 1000 g/cm$^2$. This component was named long flying component.

A few years later, the effect was confirmed in the works  cooperation {\it PAMIR} \cite{pam_abs}. In deep lead x-ray emulsion chambers (XREC), a penetrating component of cosmic rays was observed. At a depth of more than 50 cm of  lead (70 c.u.)   an absorption of hadrons with energies $E_h^{\gamma}\geq6.3$  TeV was becoming slower. The absorption length of hadrons in XREC  $\lambda_{abs}$ was changing from $200\pm5$ g/cm$^2$ to $340\pm80$ g/cm$^2$.

Both effects were explained by the possible increase in the birth cross-section of the charmed particles beyond accelerator  energies \cite{dremin}.

The experimental data  presented here confirm the existence of the penetrating component, but its interpretation is different.
First of all, this distinction is due to the fact that the effect is observed not in the lead absorber, but in the atmosphere. The hypothesis leading charm cannot explain the decrease in the absorption of the cascade in the light material (air).  In addition we also found  the energy region in which the penetrating component appears. A new component arise threshold way at an energy  3 PeV  and vanishes at energy about 100 PeV. Strange, as it may seem, this  limited energy interval  3--100 PeV  coincides with the region of restructuring of the EAS spectrum, which is occur between its two breaks.

\begin{figure}[t]
\centering
\includegraphics[width=7cm]{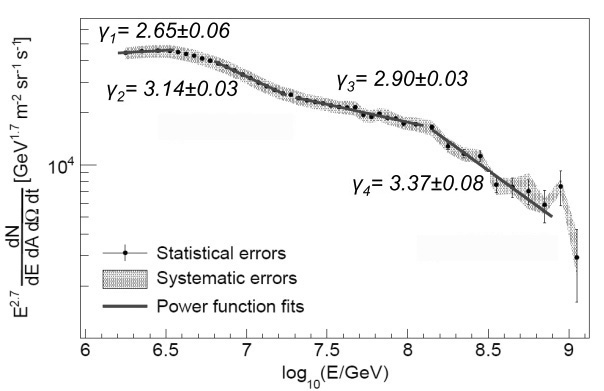}
\caption{EAS spectrum of the IceTop experiment. The numbers indicate the slope of the CR spectrum. The CR spectrum is multiplied by \texorpdfstring{$E_0^{2.7}$}{Lg}. (The figure is given with the permission of IceCube collaboration.)}
\label{f:10}
\end{figure}

The first break at $\sim3$ PeV energy was detected more than fifty years ago \cite{christ}, the second break at $\sim100$ PeV energy was obtained  in a number of experiments quite recently. The EAS spectrum received in  \cite{icetop} shown in figure \ref{f:10}.

The figure \ref{f:10} shows detailed data on the change in the slope of the EAS spectrum in the knee region. The most important feature is the presence of two breaks at energies $3\cdot10^{15}$ and $\sim10^{17}$ eV. It seems tempting to associate the energy region of 3--100 PeV between the two breaks of the EAS spectrum  with the variation of the CR mass composition. The  change in the slope of the nuclear spectra from protons to iron nuclei at the same magnetic rigidity $R=pc/Ze\simeq3$  PV just fits between two breaks in the  total particles  spectrum. However, due to the  scaling violation in the same area, the situation looks more complicated.

\section{ The {\it HADRON} installation.}

\begin{figure}[t]
\begin{minipage}[t]{0.45\linewidth}
\vspace{-6cm}
\includegraphics[width=6cm,height=6cm]{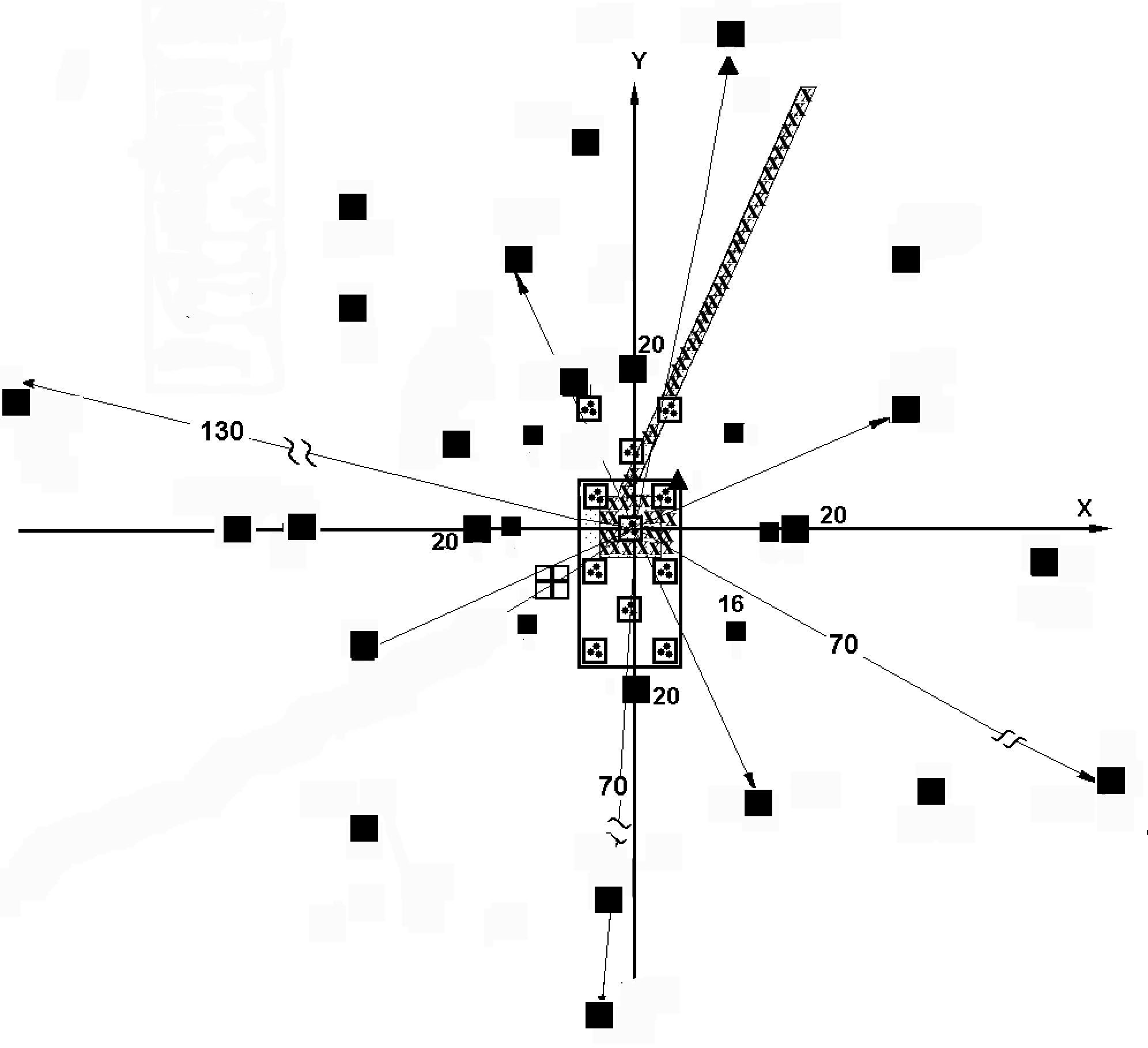}
\caption{Top view of the HADRON installation. Large and small full squares relate to the ground level scintillators \texorpdfstring{$1$}{Lg} and \texorpdfstring{$0.25 m^2$}{Lg}, respectively.   The shaded area shows location of the muon detectors in the underground laboratory.}
\label{adron_1}
\end{minipage}
\hspace{5mm}
\begin{minipage}[t]{0.45\linewidth}
\includegraphics[width=7cm,height=6cm]{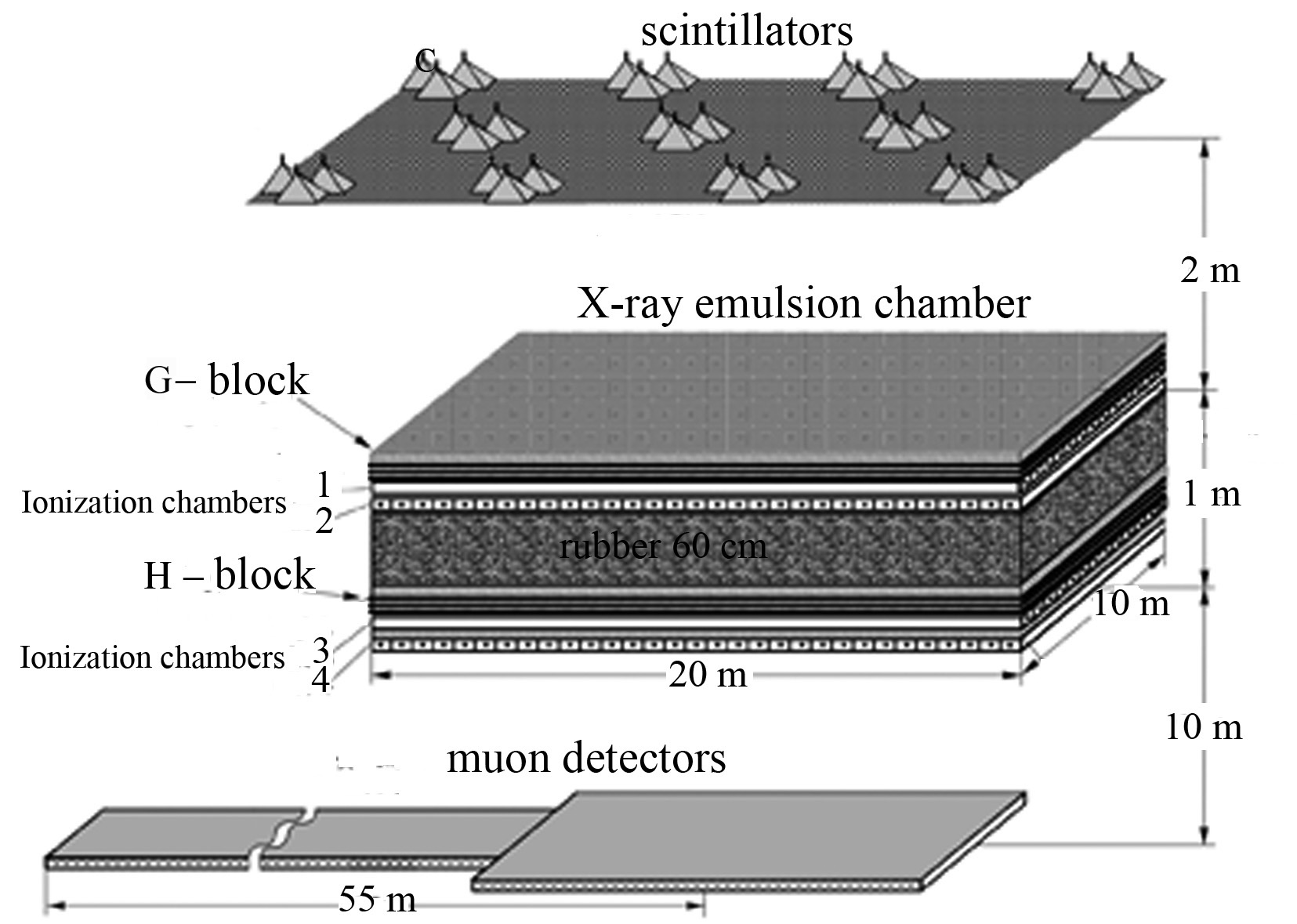}
\caption{The  central part of the HADRON installation consisting of (from top to bottom) ground-based scintillation detectors, XREC with ionization chambers and the underground muon detector. }
\label{adron_2}
\end{minipage}
\end{figure}

The XREC are unique  detectors of EAS cores. However, the lack of the information on the primary EAS energy significantly reduces the efficiency of the XREC method. Data on the CR primary energy are usually obtained from the EAS electromagnetic component.  Hybrid experiments combine the operation EAS electronic detectors with  film detectors of an XREC.  The hybrid {\it HADRON}  experiment is based on both Tien-Shan  studies of the EAS \cite{trudi109} and XREC technique developed in the frames of the {\it PAMIR} experiment \cite{trudi154}.

The association of the XREK and EAS methods is possible only if EAS cores include hadrons with energies higher few TeV. This implies the exposure of the installation at the mountains level, high enough in the atmosphere, to ensure good  efficiency of XREC usage. In addition,  Tien-Shan altitude (3330 m.a.s.l.) is close to the maximum of the shower development in the atmosphere. This reduces the dependence of data obtained on the primary nucleus mass and a chosen hadron interaction model.

The layout of scintillation counters and muon counters of the Tien-Shan EAS array  is shown in figure \ref{adron_1}. The numbers in the figure indicate the distances of the detectors from the center of the installation in meters. The central part of the installation is shown  in more detail in figure \ref{adron_2}.

The EAS array  contains  114 surface scintillation detectors of 0.25 and 1 $m^2$ in aria. The scintillators are shown as full and empty squares in figure. Large and  small full squares correspond to detectors of $1$ and $0.25 m^2$, respectively. The central part of array also includes 11 scintillator stations (empty squares) which consists of three detectors of $0.25 m^2$ each.  The scintillation detectors are placed circumferentially at radii of 16, 20, 40, 55 and 70 m. The EAS array was used to measure the primary EAS energies  and EAS arrival angles $\vartheta_x,\vartheta_y$

The area of  the muon counter hodoscope is $S=55 m^2$. It is placed at a depth of 20 m of water equivalent in the underground ($E_{\mu}>5$ GeV). The hodoscope is disposed in part under the center of the  array  and in the 50 m length tunnel  (shaded area in figure \ref{adron_1}).

The central part of array includes the XREC of 162 m$^2$ in area. The X-ray films in an XREC  detect  $\gamma$-ray and electron cascades (further, only $\gamma$-ray) developed in the lead.  Such $\gamma$-rays with energies above 2 TeV are generated mainly in $\pi^0$ decays $\pi^0\rightarrow2\gamma$. Therefore, spectra of $\gamma$-rays reflect spectra of hadrons in the EAS cores.  The high energy hadrons  on the level detection are produced for the most part by light primary nuclei because of  disintegration heavy nuclei in the upper atmosphere. As a result  XREC select EASs  formed  mostly by protons and He nuclei.

Four layers of cris-cross ionization chambers (burst detector --- BD) were placed pairwise under G- and H-blocks of XREC. The BD-installation determined the axis of the EAS  with high precision and were used to associate an events in XREC with the EAS.

\section{The XREC characteristics. }

First time the technique of XREC was proposed in the works  \cite{rapoport} and soon used for the creation of big x-ray emulsion chambers.
The rapid development of this technique in CR is associated with the relative simplicity of  XREC creating with an area of hundreds square meters. The method has a unique set of properties: it allows to determine the angles of arrival of particles, has a spatial resolution of $\sim{100}$ m, which allows to measure the individual energy of each cascade if it is higher than $\sim{2}$ TeV. This case, the spots of darkening , which  are created  on x-ray film by the cascades, can be detected visually. In addition, method for determining the energy of particles by spots photometry are much easier then counting the number of particles in a nuclear emulsions. These properties of x-ray films allowed them to displace nuclear emulsions from ground-based CR experiments.

\subsection{The \texorpdfstring{$\gamma$}{Lg}-quanta energies.}

The spot size ( $\sim{100}$ mcm) and its darkening D on the film depends  the energy of the primary $\gamma$-quantum. The energy  is determined  by  photometry of the spot  by  diaphragms of different radii: 48, 84, 140 and 290 mcm, allowing to obtain the values of the average darkening for several sections of the spot \cite{trudi154}. These data  compared to darkening profiles calculated according to the theory of electron-photon cascades \cite{ivanenko}. The relationship between the darkening and the number of electrons n, which have created it,  was established by means of  the empirical formula  $D(n)=D_{\infty}[1-exp(-ns)]$ (Smorodin). Here $D_{\infty}=4$ is the maximum darkening for the single emulsion layer and $s\simeq(3.25\pm0.13)\text{mcm}^{2}$ is the effective area of silver granules in the films.

The measured value of $D_{mes}$ for a number of methodical reasons can differ from the theoretical density of darkening. Therefore true
value of $D_{true}$ was obtained after taking into account the corrections on instability of film development conditions, non-standard
photometer work,  regression of the hidden image, the presence of gaps in a lead absorber and other. The main sources of distortion are discussed below.

\begin{figure}[t]
\centering\includegraphics[width=7cm]{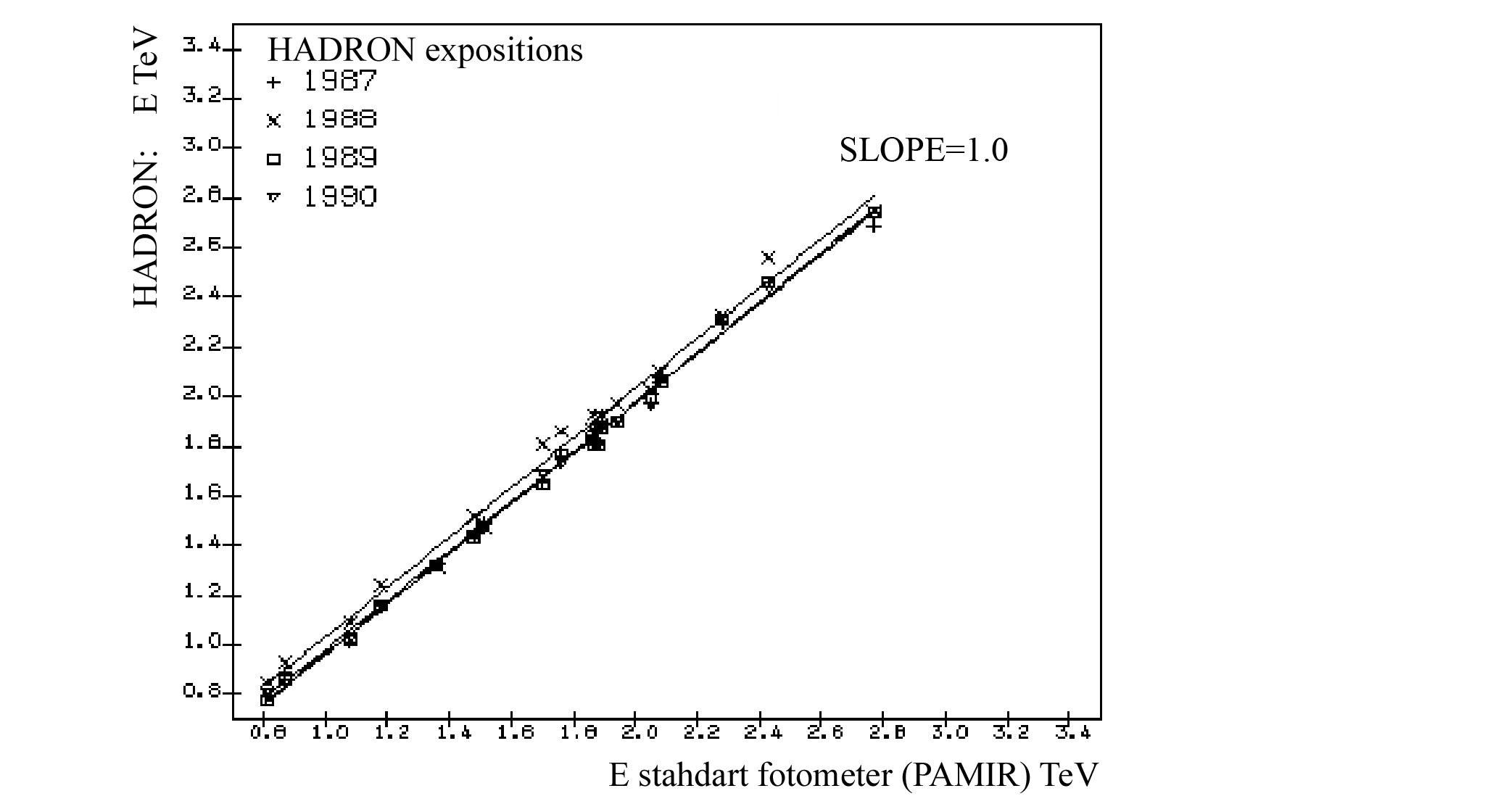}
\caption{Calibration of photometers of the HADRON experiment according to the standard photometer of the PAMIR experiment.}
\label{dolg}
\end{figure}

\begin{itemize}

\item {\bf Diffused light.} The gradient of the darkening density in the spot depends on its size, which leads to a  dependence of the scattered light on the primary energy $\gamma$-rays. Therefore, a change in the aperture in the optical channel can change the measurement conditions, lead to a different value of the measured light and change the values D.

Measurements  densities of darkening with different apertures of lenses allowed to take into account the role of scattered light. It is found that when the density $D\leq3$ the density was overstated by 5--7$\%$. This amendment was taken into account in the  measurements on the so-called standard photometer. The readings of photometers of the {\it HADRON} experiment were given to the readings of the standard photometer by means of the control  spots of darkening. The calibration accuracies in different expositions are shown in figure \ref{dolg}. The readings of the photometers is consistent throughout the energy range, the slope of the dependence of $E_{hadron}(E_{stand})$ is not different from unity. Some increase in the measured energy in  exposition of 1988 year is constantly and  does not exceed $10\%$.

\item {\bf Determination of the cascade center.} Another source of measurement error is the wrong in the determination of the axis of the cascade, connected mainly with the graininess of the image. The maximum darkening (center) was determined by the minimum of the  PMT photo-current, which was  used as the photo detector.  The error  into value D  was  $10\%$ for $E_{\gamma}=$3--5 TeV and 5--6$\%$ for $E_{\gamma}\approx{20}$  TeV.

\item {\bf Photo image regression and film development.} Control over the development of films was carried out using seven standard  markers $D_m$ of different density, which were applied to each film by radioactive sources before films development. The dispersion of $D_m$  with respect to their average value was $\sim10\%$ for a group of films, which were developed at the same time. The  correction of  the  darkening  D determined by means of the ratio  $D'_{meas}=D_{meas}\cdot{\bar{D}_m}/D_m$.

The regression of the photographic image on the film was taken into account on average by markers which were applied before and after the exposure of films. The following empirical expression for the regression coefficient was obtained from the measurements $K_{reg}=1.064-0.0267\cdot D'_{\text{meas}}$ for $D'_\text{meas}<2.4$ and =1 for
$D'_\text{meas}\geq 2.4$, where $D''_{\text{meas}}=K_{reg}\cdot D'_{\text{meas}}$.

\item {\bf The angle of incidence.} The correction for the angle of incidence EAS were introduced by the coefficient $K_\theta$, which  convert  the measured density $D_{meas}$ to conditions of the   normal fall. The correction is small and can be neglected for angles $\theta=0$--$35^{\circ}$.
\end{itemize}

The energy  dependence of the $E_{\gamma}$ errors  can be determined by means of the  fluctuations $E^{\iota}_{\gamma}$ which were received for three depths  of  XREC. The real  errors were obtained normalizing its by data  of absolute energy calibration. For this purpose, a calibration chamber with a carbon target located above a multi-layer Pb chamber at a height of 5.8 m was  created in the Pamir. The decay events $\pi^0\rightarrow2\gamma$ were selected and the $\pi^0$-meson mass  was determined by measuring the opening angle and energies of $\gamma$-quanta. The error amounted to the value of $\sigma(E_{\gamma})/E_{\gamma}\simeq 0.4$ for energies $\gamma$-quanta 2--4 TeV.

\begin{figure}[t]
\begin{center}
\begin{minipage}[t]{0.45\linewidth}
\includegraphics[width=6cm,height=6cm]{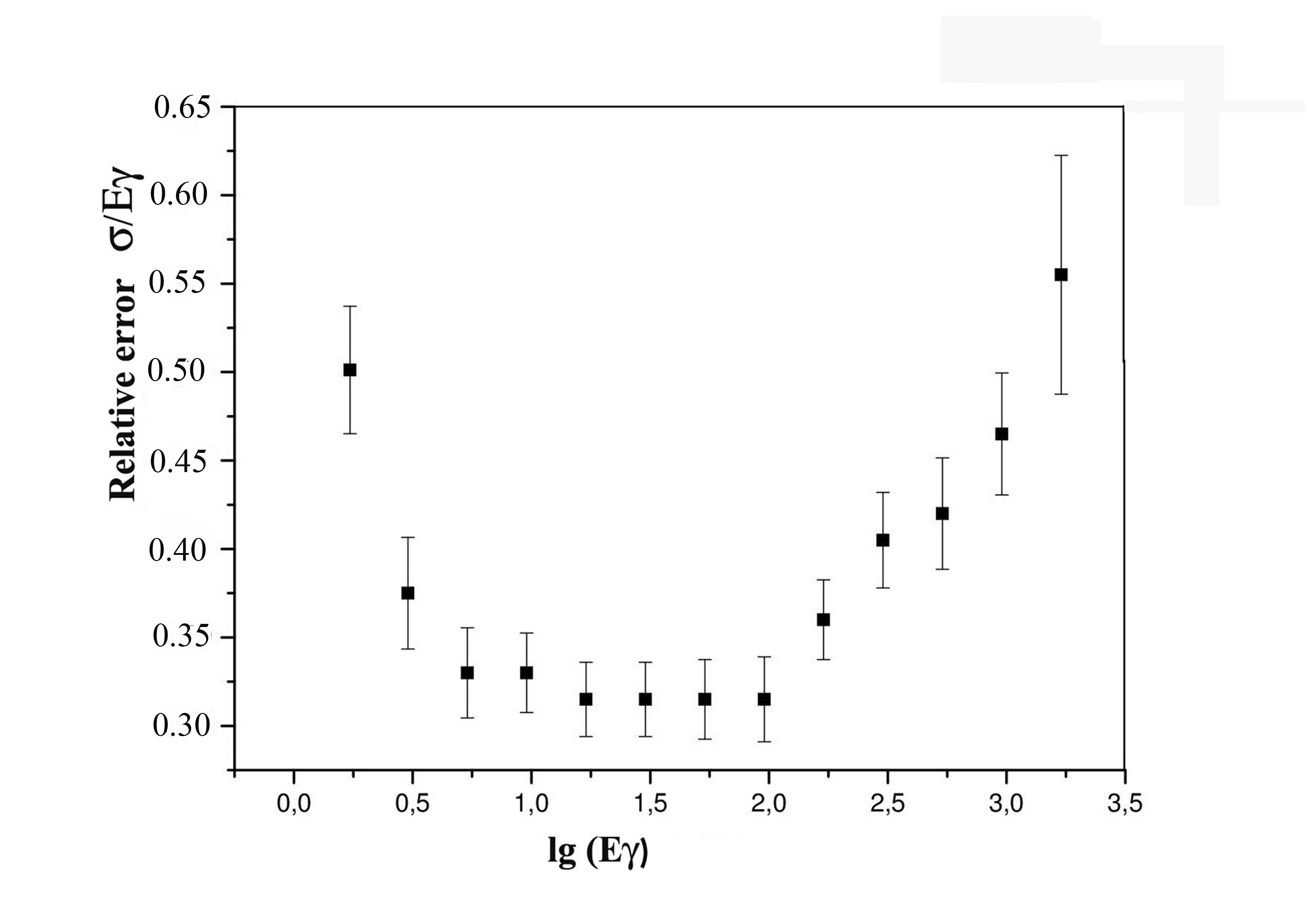}
\caption{The relative accuracy of the energy determination of \texorpdfstring{$\gamma$}{Lg}-rays in XREC. }
\label{f:6_l}
\end{minipage}
\hspace{5mm}
\begin{minipage}[t]{0.45\linewidth}
\includegraphics[width=6cm,height=5cm]{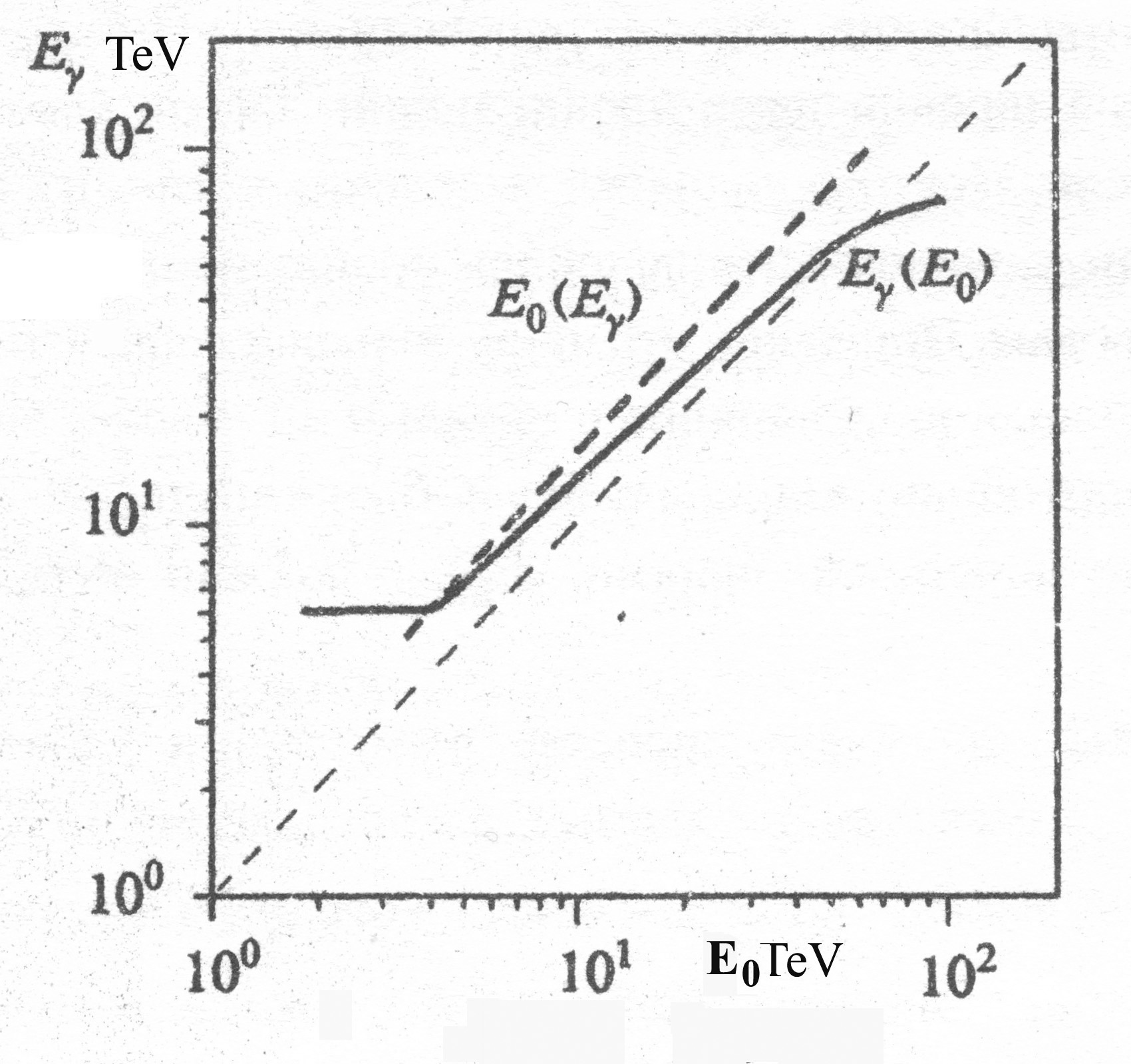}
\caption{Dependence of the measured energy \texorpdfstring{$E_{\gamma}$}{Lg} on the true \texorpdfstring{$E_0$}{Lg} and Vice versa. Monte-Carlo calculations. }
\label{f:6_r}
\end{minipage}
\end{center}
\end{figure}

The normalized errors  are shown in figure \ref{f:6_l}. In the region of $E_{\gamma}$ spectra $E_{\gamma}=$6--100 TeV the error are almost constant and not exceed $\sigma(E_{\gamma})/E_{\gamma}\simeq0.35$. When constructing the energy spectra of $\gamma$-rays in each energy interval, averaging over many quanta occurs, so the average errors in this case is several times less.

Figure \ref{f:6_r} illustrates  the effect of overlapping closely spaced cascades on the energy of  $\gamma$-quantum . The dependence is obtained by calculation. The recovered energy $E_{\gamma}$ is systematically higher at $E_0<70$ TeV. The maximum distortion is observed in 1.5 times on the threshold of registration of $\gamma$-quanta. In the experiment, the mutual influence of cascades was eliminated programmatically.

In determining the energy of $\gamma$-quanta, all these amendments were taken into account.

\subsection{The arrival angles in an XREC.}

\begin{figure}[t]
\centering\includegraphics[width=7cm]{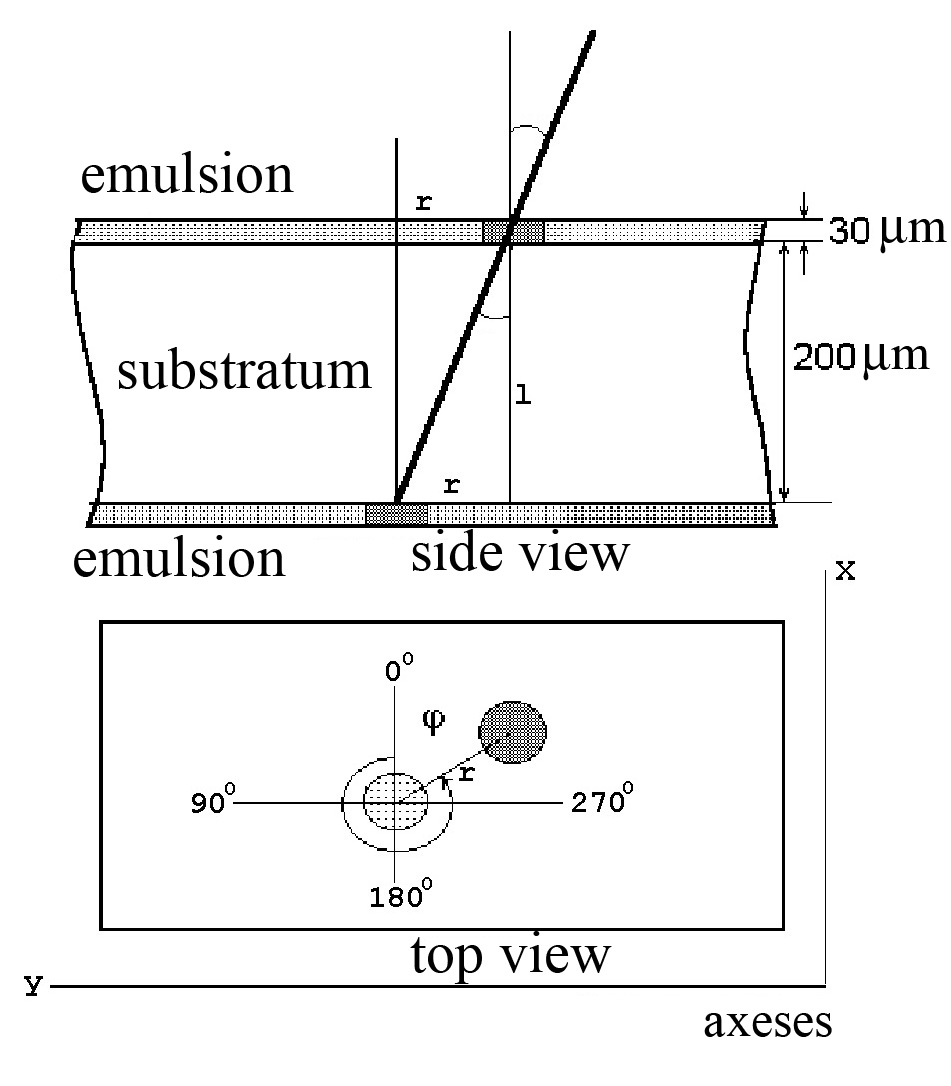}
\caption{Angular measurements in XREC films.}
\label{angl}
\end{figure}

The structure of the X-ray films is shown in figure \ref{angl}. The transparent base has a thickness of 200 mcm on both sides of which the emulsion is applied.   The relative shift of dark spots in these layers allows you to define the arrival angles of the cascades $\theta$, $\varphi$  or $\theta_x$, $\theta_y$ .  The process of measuring angles in XREC is shown in the same figure. Cascades with equal angles  within the radius of 15 cm are considered to be genetically related and are combined into  $\gamma-$families.

The zenith angle of the cascade is determined by the ratio $tg\theta=L/D$, where L is the distance between the darkening spots  on the upper and the lower layers of the emulsion and D the thickness of the substrate.  The azimuthal angle $\varphi$ equals the angle between  coordinate axis 'x' and a straight line in the plane of the film passing through both darkening spots. The $\varphi$ is measured from the x-axis in a counterclockwise direction if see  in the direction "to the source" of the cascade.

The measurement accuracy of the  $\theta$ angle is about $\sigma_{\theta}=6^{\circ}$. When events in XREC are combining with  EAS,  the $\theta_x$, $\theta_y$  angles are used instead  angles $\theta$ and $\varphi$. The measurement accuracy of the corresponding angles is $\sigma_{\theta_x}=\sigma_{\theta_y}\simeq8^{\circ}$. For $\gamma$-families  this value is reduced in $\sqrt{n_{\gamma}}$ times, where $n_{\gamma}$ is the $\gamma$-quanta multiplicity in this family.

\section{The EAS characteristics.}

The registration of events  was started by two triggers: scintillation and burst. The scintillation trigger provided registration of EAS with an axis inside a circle with a radius of 70 meters and a threshold of $N_e>3 \cdot10^5$. The burst master was produced if in any of the BD ionization chambers the pulse exceeded the threshold corresponding to the passage through the chamber  $1.3\cdot10^4$ particles (energy 2--3 TeV per channel).

\subsection{The EAS angles.}

The direction of EAS arrival  was determined by a standard device {\it Chronotron} . It consisted   four temporal scintillation stations located at distances of 20 m from the center of the setup.
The delays $\Delta\tau_{x}$ and $\Delta\tau_{y}$  measured by the pairs of  detectors 1-3 and 2-4.  Measurements of times were conducted across discrete intervals
$\Delta\tau=8$ ns in the range $t=0\div{248}$ ns.

The accuracy of the angles reconstruction  was determined by simulation  the zenithal angular distribution in the form $f(\theta)\sim \cos^6\theta\sin\theta$ and uniform azimuthal distribution.

The  contribution of  discreteness the time measurement  on these errors is relatively small and is  $\sigma_{\theta}\simeq1.3^{\circ}$ and $\sigma_{\varphi}\simeq3.6^{\circ}$. The main contribution to errors give  EAS front fluctuations. The final  accuracies  of the angles determination were $\sigma_{\theta_x}=\sigma_{\theta_y}\simeq9^{\circ}$.

\subsection{The total number of electrons  \texorpdfstring{$N_e$}{Lg} and the age parameter 's'.}

The energy of the shower is determined by model calculations using the total number of electrons $N_e$ in the shower at the observation level. The total number of electrons was determined by integration of the lateral distribution function (LDF)  of charged particles obtained from the data of scintillation detectors.
To receive LDF the SC data were  approximated  by the Nishimura-Kamata-Greisen (NKG) functions \cite{NKG}:

\begin{equation}
f_{NKG}(r)=C(s)\cdot(\frac{r}{r_m})^{s-2.0}\cdot(1-\frac{r}{r_m})^{s-4.5},
\label{fl:1}
\end{equation}
where s-age parameter, $r_m=125$ m --- radius Molier and C(s) is determined by the expression

\begin{equation}
C(s)=\frac{1}{2\pi}\frac{\Gamma(4.5-s)}{\Gamma(s)\Gamma(4.5-2s)}
\label{fl:2}
\end{equation}

The densities of electrons in the EAS are associated with the NKG function by the following relation:
\begin{equation}
\rho_{e^-}=\frac{N_e}{r_m^2}f_{NKG}.
\label{fl:3}
\end{equation}

\begin{figure}[t]
\begin{minipage}[t]{0.45\linewidth}
\includegraphics[width=6cm,height=6cm]{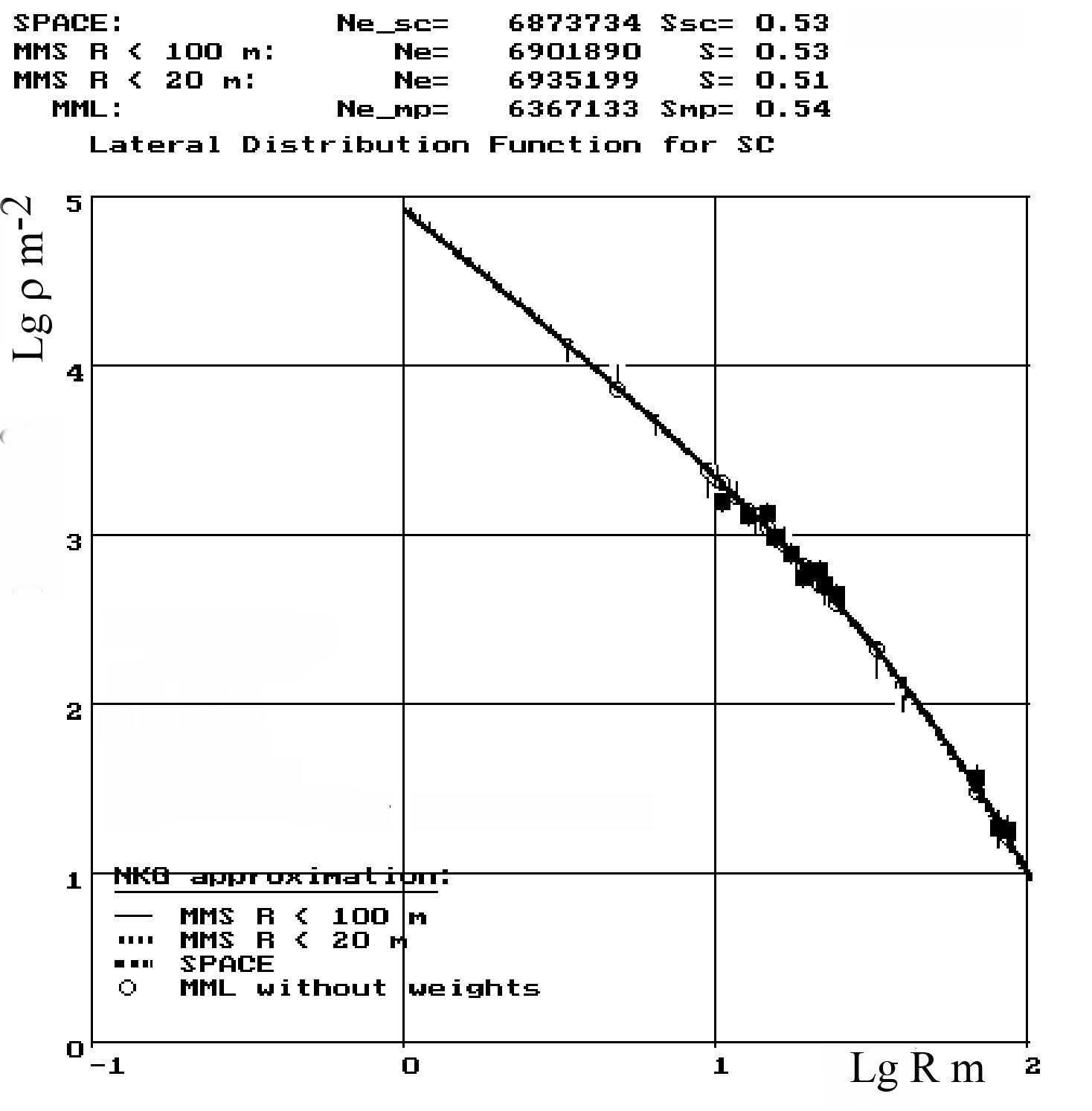}
\caption{ An example of  EAS in which all four approximations are practically the same (see text).}
\label{ldf_l}
\end{minipage}
\hspace{5mm}
\begin{minipage}[t]{0.45\linewidth}
\includegraphics[width=7cm,height=6cm]{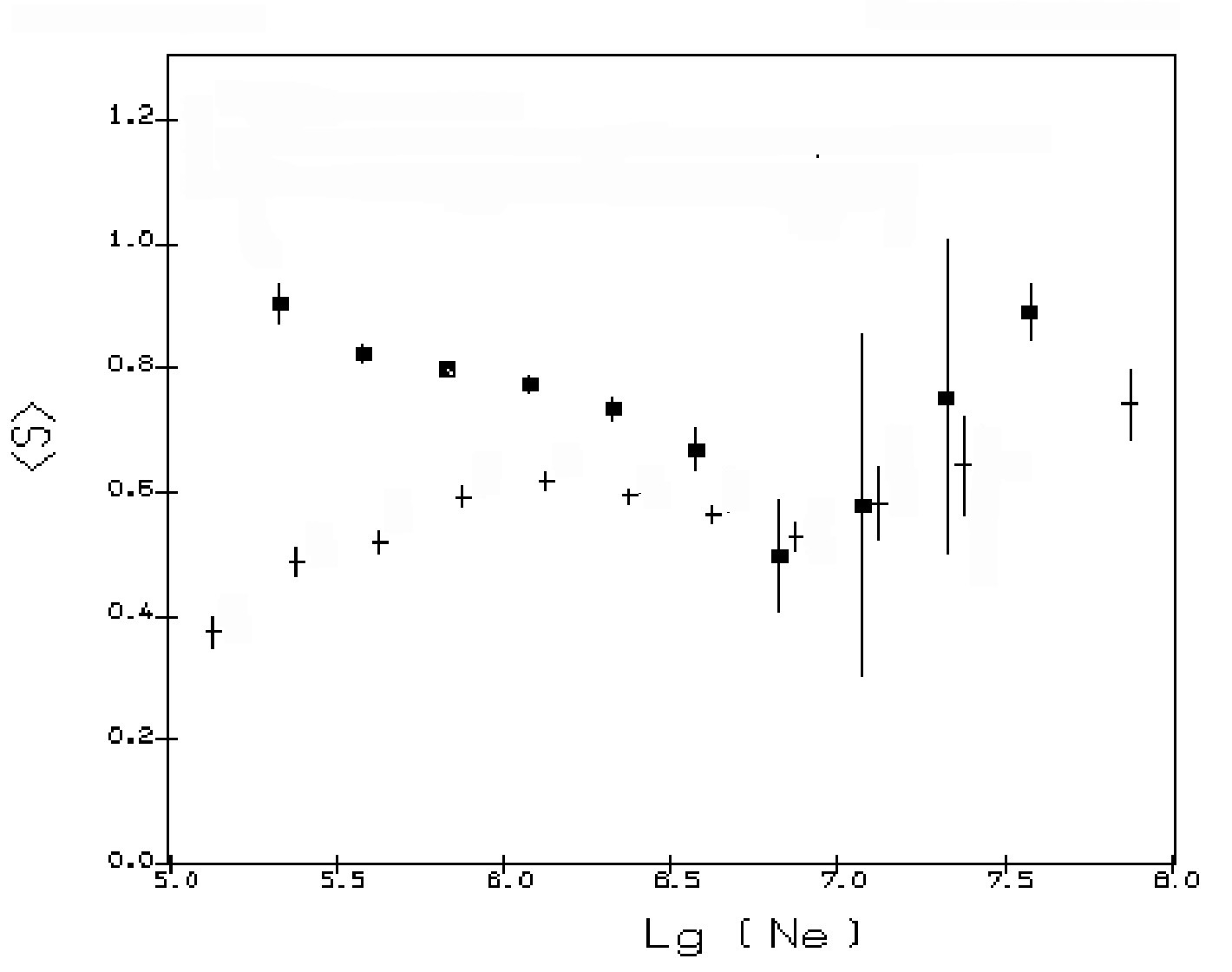}
\caption{The \texorpdfstring{$N_e$}{Lg} dependence of the EAS transverse age 's'   for all EAS and EAS with \texorpdfstring{$\gamma$}{Lg}-rays:
\texorpdfstring{$\blacksquare$}{Lg} --- EASs, \texorpdfstring{$+$}{Lg} --- EAS with \texorpdfstring{$\gamma$}{Lg}-rays. }
\label{ldf_r}
\end{minipage}
\end{figure}

The LDF approximation is reduced finding the age parameter 's' and normalization of theoretical densities on the experimental values. We were used three main methods for determining 's': the maximum likelihood method, the least squares method and the least squares method with weights. In addition, in the least squares method, the value of the parameter 's' was determined by detectors located at distances  $R_i<20$ m from the shower axis. The values of 's' and the corresponding LDF are slightly different in these methods. It leads to a difference in the values of the total number of electrons $N_e$. In  70 per cent of cases the approximations are virtually identical and the difference in $N_e$ is 5--6$\%$. An example of such event is shown in figure \ref{ldf_l} where all four approximations are given. For all events, the error of $N_e$ did not exceed value $\sigma=0.12$.  The main contribution to the error is given by the approximation option at $r<20$ m which in the future is not used.  Ultimately   the $N_e$ error   lies in the range 5--10$\%$.

In general case the EAS  parameters  have been looking for by the maximum likelihood method (Pavluchenko Space  algorithm). To find the shower axis according to SC detectors the functional was constructed with  global minimum in the region of axis.  To receive it, the parameters of the LDF slope  'a' were  calculated  for any two points of registration  in the assumption that $\rho(r)={r^{-a}}$.  The dispersion of the set  weighted values  'a' (pseudo age) were considered as the functional  of  zero approximation. The axis coordinates x,y were calculated iteratively, and $N_e$ was  calculated once at the end of iterations. Monte Carlo calculations had shown that the accuracy of the EAS axis determination was $\Delta{r}\simeq0.9$ m and the precisions of determining the parameters 's' and value $N_e$ were $\Delta{s}\simeq0.05$ and
$\Delta{N_e}/N_e\simeq0.1$, respectively (algorithm Space).

If the axis of the shower falls into the burst detector and, accordingly, in XREC, its position can be determined by the ionization chambers with an accuracy  0.25 m, which is four times more accurate than in the Space algorithm.  The $N_e$ error mainly depends on the accuracy of the EAS axis definition, hence for showers falling into XREC it should be really less than $10\%$.

The figure \ref{ldf_r} shows the $N_e$ dependence of the EAS age parameter 's'   for all EAS and EAS combined with $\gamma$-rays in XREC. The dependence has a non monotonic character with a minimum at $N_e\simeq10^7$. A decrease of 's' in the region $N_e<10^7$ means that the LDF of electrons in the EAS become steeper. This can not be explained by the increase in energy, because in the region  $N_e>10^7$ there is an inverse relationship, parameter 's' increases for both EAS components. The sharp increase in the dispersion in this area is noteworthy. This may mean that two EAS groups are in this region  with significantly different values of the parameter 's'.

The presence of showers with small values of the 's' parameter caused the need for the introduction of weights  into the least squares method. The increase in the steepness of the LDF led to an abnormally large contribution to the standard quadratic form of Q detectors with maximum particle densities near the shower axis. The introduction of weights reduced the relative contribution of these members and allowed more stable estimates of the age parameter. The parameter 's' in that case was determined by the minimum of the following quadratic form (algorithm Q100):

\begin{equation}
F=\sum\limits_{\iota=1}^{k}\frac{(\rho_{\iota}^{exp}-\rho_{\iota}^{NKG})^2}{\rho_{\iota}^{NKG}\sigma_{\iota}},
\label{fl:4.39}
\end{equation}
where $\rho_{\iota}^{exp}$ and $\rho_{\iota}^{NKG}$ are the experimental and theoretical electron densities, respectively, and $\sigma_{\iota}$ are the experimental  errors of density in points 'i'.

Here, unlike the standard  quadratic form with $\sigma_{\iota}^2$ in the denominator, one value of $\sigma_{\iota}$ is replaced by $\rho_{\iota}^{NKG}$. This option can be considered as the least squares method with weights $1/\sigma_{\iota}$, since  there is a ratio  $D(\rho_{\iota})=\sigma_{\iota}^2=\rho_{\iota}^{NKG}$, which must  performed for  distribution of the electrone densities  according to Poisson's law.

In the future, the values $N_e$ obtained by the Q100 algorithm are used.

The primary energies of an EAS were determined by the formula  $E_0=15.1\cdot{N_e}^{0.84}$ PeV. Sometimes a simpler formula is used: $E_0\simeq{N_e\cdot\overline{E_e}}=N_e\cdot{2}$ GeV.  The average difference between calculations using these formulas did not exceeds
$\sim{20\%}$.

\section{The procedure of combining EAS and XREC events.}

The combination of EAS data and XREC events provides a unique possibility to study the EAS cores. A problem connected with the lack of time selection of events in XREC exist in this method. Events accumulated on the x-ray films during a year should be associated with an EAS after end of exposition.

The combination of XREC events  and EASs  are usually made by comparing arrival angles $\vartheta$, $\varphi$, and also coordinates of both the EAS axis and center of mass for the $\gamma$-family. In contrast to this simple method, our procedure was more complicated.   The thorough  statistical method was developed by the author which  was shown the necessity of more stronger  criteria for selection of true candidates among EAS. Next additional criteria were used to provide a low  background of incorrectly connected events:

\begin{enumerate}
  \item The EAS array should operate not less than 90$\%$ of calendar time during XREC exposition (one year).
  \item The position of EAS axis should be determined with an accuracy not worse than 25 cm.
  \item The additional  selection of EAS candidates has to be done. For this purpose, the EAS with local bursts in BD below 2 TeV were excluded from the consideration.
 \end{enumerate}

Two first  requirements considerably hampered the action of the {\it HADRON} array. The subsystem of data accumulation  was doubled to support 90$\%$ efficiency in the case of violation the condition  operation. The design of electronics allowed to carry out fast replacement of any channel if necessary. In order to avoid the galvanic coupling in the array electronic  circuits, special devices were used to transform electric signals to photon ones. This was made to protect the array from influence of thunderstorms  and induced electric noise.

The last condition uses the concept of local burst. To estimate the number of particles in the local area $0.25\times0.25 m^2$ of the the ionization chamber with the maximal burst the following procedure was used. Given the sharp dependence of LDF of ionization  on the distance in BD the readings of neighboring chambers were subtracted from the chamber readings with the maximum burst. In this case, the contribution to the ionization of the distances greater than 0.25 m was almost completely excluded, and the ionization in the maximum  has been decreasing by no more than  15--20$\%$. The formula for determining the localized bursts was as follows:
\begin{equation}
E_{l.b.}={\max}\left\{
\begin{array}{lcc}
[N_{max}-0.5(N_{max-1}+N_{max+1})]\times{2.10^{-4}}\,TeV, &\cr
[N_{max}+N_{max+1}-N_{max-1}-N_{max+2}]\times{10^{-4}}\,TeV, &\cr
[N_{max}+N_{max-1}-N_{max-2}-N_{max+1}]\times{10^{-4}}\,TeV, &
\end{array}\right.
\end{equation}
where N is the number of particles in the chamber.

The events in XREC  at energies $\Sigma{E_{\gamma}}>10$ TeV were selected to associate them with an EAS. As was mentioned above, showers with bursts in BD lower than 2 TeV were excluded from the analyses.  The check showed that this procedure does not distort features of  combined EASs. It was found that depending on  the energy  $\Sigma{E_{\gamma}}$  from 25 to 1 candidates of EASs per one $\gamma$-family remained after this selection.

For remaining EASs, the usual procedure based on arrival coordinates and angles was applied.

The  procedure of association was carried out using the Neyman-Pearson  criterion for the three parameters:  $\overline{x}=(\theta_{x},\theta_{y},R_{{\text{eas}},{\gamma}})$ of EASs and
$\gamma$-families. The distribution of errors for the variable $\varphi$ has a non-Gaussian shape , therefore, the equivalent pair of variables $\theta_{x},\theta_{y}$ was used.

The same expressions can be written for all parameters used in combining:
\begin{align}
  l_{\vartheta_x}^{ij}=\frac
{P(\vartheta_{x})}
{2\pi\sigma_{\vartheta_{x1}}\sigma_{\vartheta_{x2}} P_i(\vartheta_{x1})P_j(\vartheta_{x2})}
 \exp(-\frac
{(\Delta{\vartheta_x})^2}
{2(\sigma_{\vartheta_{x1}}^2+\sigma_{\vartheta_{x2}}^2)}
), \\
l_{\vartheta_y}^{ij}=\frac
{P(\vartheta_{y})}
{2\pi \sigma_{\vartheta_{y1}} \sigma_{\vartheta_{y2}} P_{i}(\vartheta_{y1})P_{j}(\vartheta_{y2})}
 \exp(-\frac
{(\Delta{\vartheta_y})^2}
{2(\sigma_{\vartheta_{y1}}^2+\sigma_{\vartheta_{y2}}^2)}
), \\
l_R^{ij}=\frac
{\Delta S}
{S_0 2\pi \sigma_{R1}\sigma_{R2}}
\exp(-\frac
{\Delta R^2}
{2(\sigma_{R1}^2+\sigma_{R2}^2)}
).
\end{align}
Here, $S_0=0.25\times0.25$ m$^2$ is the intersection area of crossed ionization chambers, which determines the accuracy of the EAS axis position,
S=162 m$^2$ is the total area of the XREC, $\Delta R$ is the distance between the center of mass for the gamma-family and  the center of mass for ionization distribution in the BD. The subscripts i,j and 1,2 refer to the pair  $EAS_i,XREC_j$ being tested.

Finally, the combining criterion is as the following:

\begin{equation}\label{H1}
 l^{j}=\max\vert _{i=1}^{n_j}(l_{\vartheta_x}^{ij}\cdot  l_{\vartheta_y}^{ij}\cdot  l_R^{ij}) \geq C_{\alpha},
\end{equation}

where the maximum is searched among all  $n_j$  candidates of $EAS_i$  for each $XREC_j$.

The value of $C_{\alpha}$ was chosen empirically.

\begin{figure}[t]
\centering\includegraphics[width=8cm]{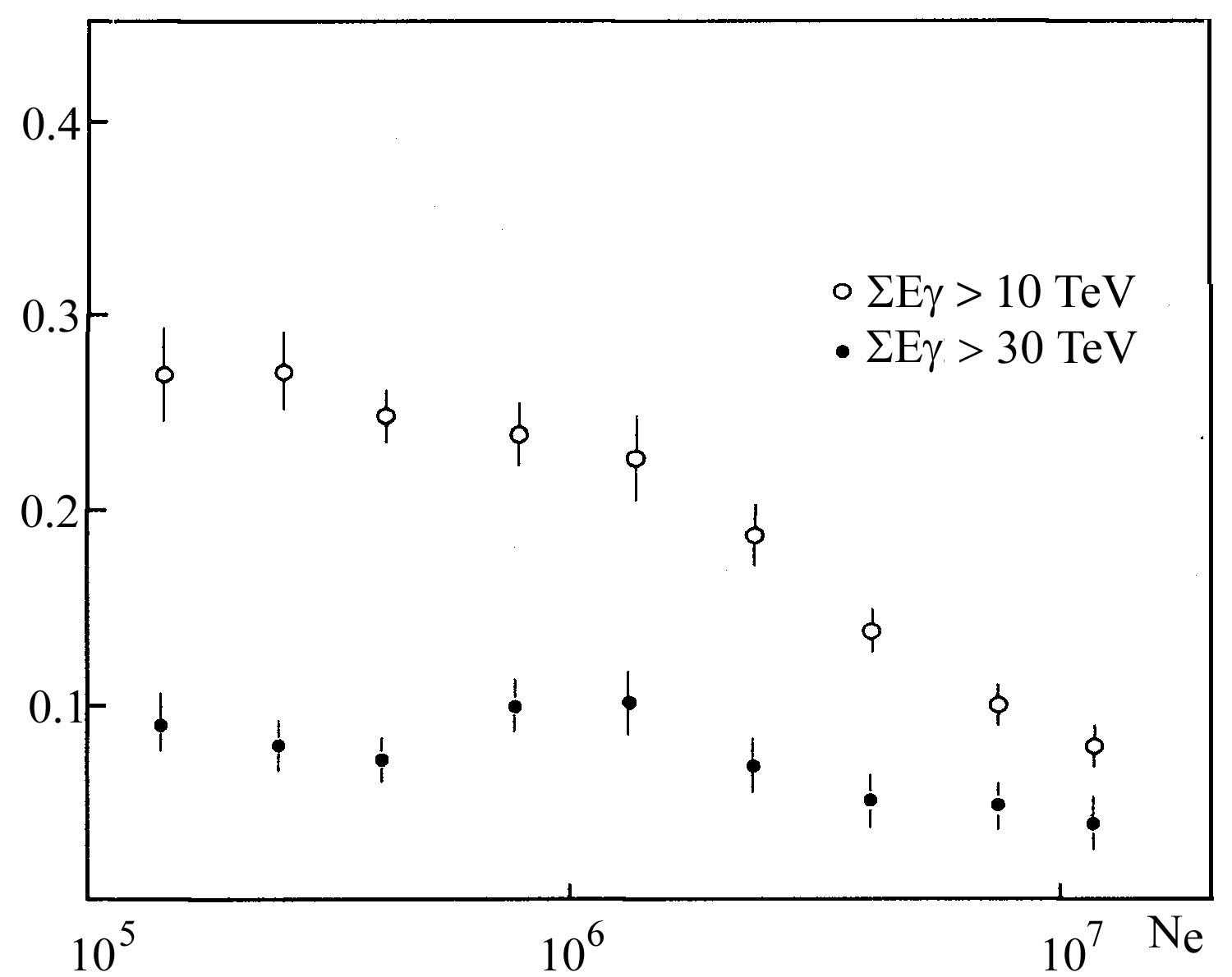}
\caption{ \texorpdfstring{$N_e$}{Lg} dependence of the share of background events among the combined  \texorpdfstring{$\gamma$}{Lg}-families with  \texorpdfstring{$\Sigma{E_ {\gamma}}\geq 10$}{Lg} and 30 TeV.}
\label{bgr}
\end{figure}

Figure \ref{bgr} shows the proportion of background events, i.e. events that are incorrectly matched, depending on $N_e$. In the  region $N_e=10^6-10^7$, the background fraction for $\gamma$-families with $\sigma E_{\gamma}\geq 10$ TeV decreases from $25$ to $10\%$. At $\sigma E_{\gamma}\geq 30$ TeV it does not exceed $10\%$ in the whole range $N_e$.

\section{The experimental results.}

The combination of $\gamma$-families with EAS was made for four XREC expositions. Their durations were 0.5 and 1 year for the first two and the next two expositions, respectively. The statistics  of the combined events amounted 1665 families with $\Sigma{E_{\gamma}}\geq10$ TeV for  the geometric factor  $ST=480$ m$^2\cdot$year.

At the beginning of the experiment we were interested in determining  of the CR mass composition by means integral spectra $>E_{\gamma}$  in the region of the primary energies $E_0>1$ PeV. The calculations were showing that the slope of the spectrum of $\gamma$-rays 'b' ($I (>E_{\gamma})\sim E_{\gamma}^{-b}$ in XREC must changes from $b\simeq1\,\text{to}\, 2$ when the kind of the primary nucleus changes from protons to Fe nuclei. We hoped that this method would provide new information on the mass composition of cosmic rays.  It was expected that in the area of the so-called knee in the CR spectrum, the nuclear composition should  become heavier and the hadron spectra may become more soft in that case.  The result was unexpected. In the local energy region of 3--100 PeV, the hadron energies increased, i.e. conversely the spectra become more hard.

Usually all spectra energies  are concluded  in the same range $E_{\gamma}=$6--100 TeV and spectra differ only in slopes and multiplicities. In order to exclude them overlapping at the figure it is convenient to introduce the  variable $x=E_{\gamma}/E_0$. Here $E_0=0.002\cdot N_e$ is the average energy for given interval $N_e$ which is the same for all  spectra have been  averaging in this interval. That case the $\gamma$-ray energy spectra have the same inclination, form, multiplicity  as the spectra $E_{\gamma}$ but only are shifted on energy axis and don't overlap each other.

\begin{figure}[t]
\begin{minipage}[t]{0.45\linewidth}
\includegraphics[width=8cm,height=6cm]{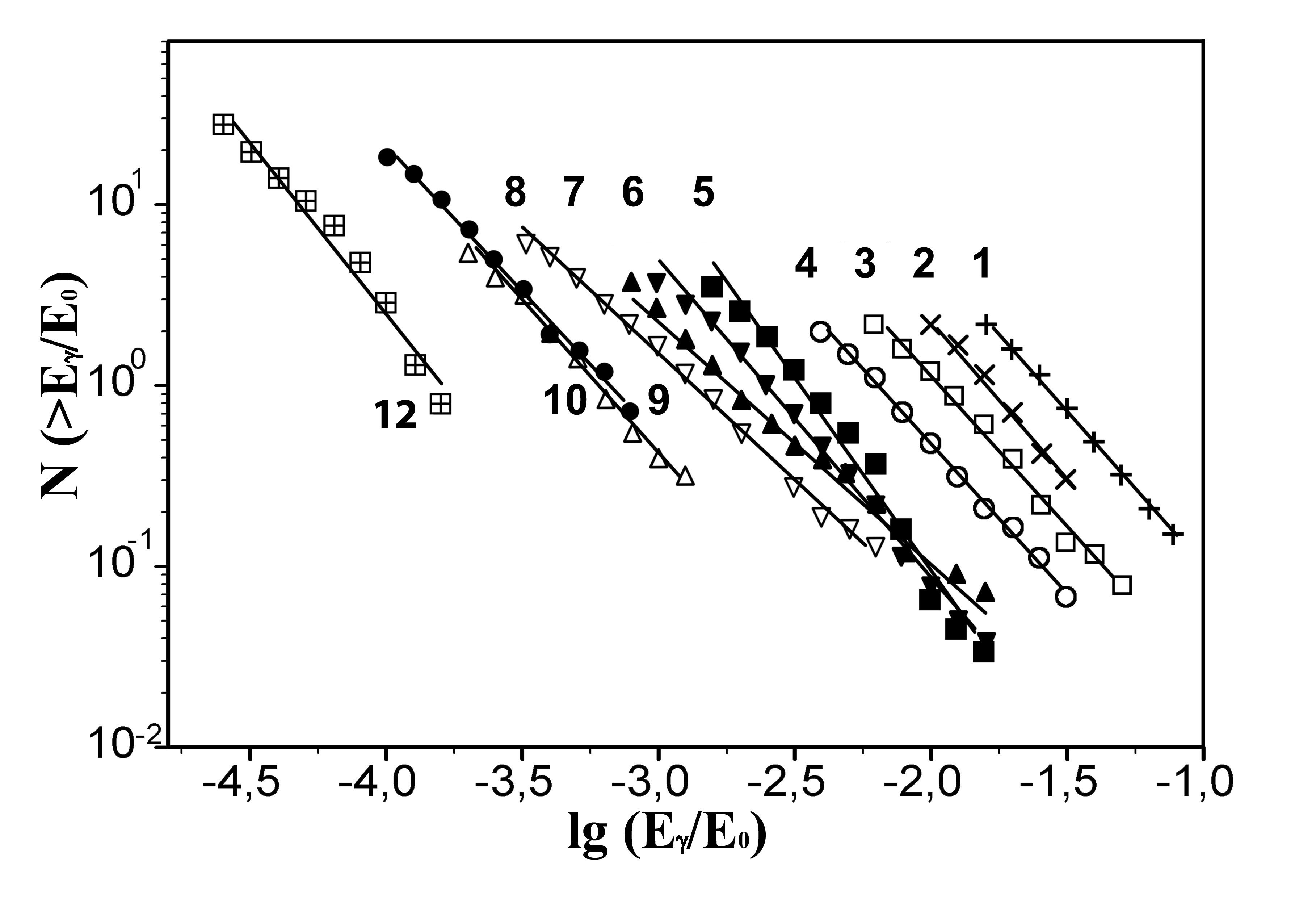}
\caption{  Normalized integral spectra of \texorpdfstring{$x=E_{\gamma}/E_0$}{Lg} in different \texorpdfstring{$N_e$}{Lg} intervals for \texorpdfstring{$\gamma$}{Lg}-families of \texorpdfstring{$\Sigma{E_{\gamma}}\geq10$}{Lg} TeV. The numbers in the figure correspond to the logarithmic interval numbers in the range
\texorpdfstring{$\lg(N_e)=4.93--7.93$}{Lg} with discreteness \texorpdfstring{$\Delta\lg(N_e)=0.25$}{Lg}.}
\label{sp_b_l}
\end{minipage}
\hfill
\begin{minipage}[t]{0.45\linewidth}
\includegraphics[width=7cm,height=6cm]{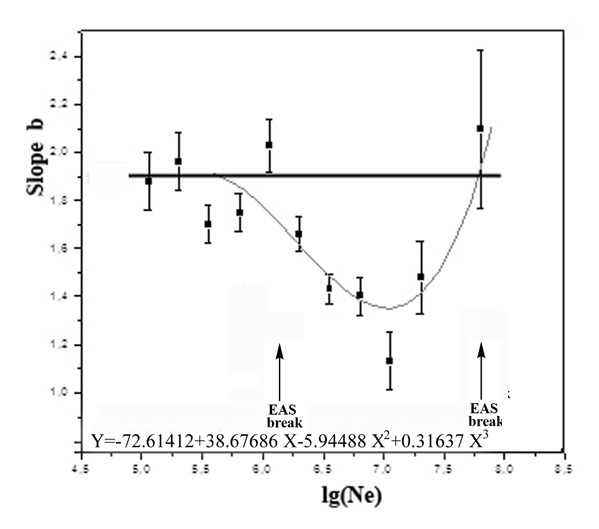}
\caption{\texorpdfstring{$N_e$}{Lg} dependence of the slope \texorpdfstring{$<b>$}{Lg} for the integral spectra \texorpdfstring{$N(\geq{E_{\gamma}})\sim E_{\gamma}^{-<b>}$}{Lg} in \texorpdfstring{$\gamma$}{Lg}-families of \texorpdfstring{$\Sigma{E_{\gamma}}\geq10$}{Lg} TeV. The horizontal line 
\texorpdfstring{$b=1.9$}{Lg} is the scaling prediction. }
\label{sp_b_r}
\end{minipage}
\end{figure}

The experimental spectra x are shown in figure \ref{sp_b_l}. The numbers in the figure correspond to ordering numbers of logarithmic intervals within the range $\lg N_e=$4.93--7.93 with discreteness $\Delta\lg N_e=0.25$. The width of the averaging interval was chosen much larger than the error $N_e$.

The spectra are shown in the energy intervals $E_{\gamma}$, where approximately they have a power form. In order to determine these intervals the differential spectra were originally constructed.  As the maxima of all these spectra were located below the energy $E_{\gamma}^{max}=6$ TeV,   all integral spectra were constructing in the higher energy range.

In addition, the spectra were normalized to the average multiplicity of $\gamma$-quanta in the families of this interval. The upper points of the spectra on the figure correspond to these values. The minimum value of $x$ in different spectra should decrease inversely proportional to  $N_e$, as far as $E_{\gamma}^{min}=6$  TeV constant for all intervals of $N_e$. The disproportional spectrum shift in the 10th energy interval, is connected namely normalizing procedure, i.e. connected with the rapid growth of $\gamma$-rays multiplicity between 9 and 10 intervals, which shift spectrum in vertical direction.

The $N_e$ dependence of the spectral indexes  'b'   are presented in figure \ref{sp_b_r}. It shows a local 'b' changing  in the range $N_e=10^6$--$10^8$ with the maximum decrease of the slope at $N_e=10^7$. This dependence corresponds to an increase in the average energy of hadrons, i.e. means the appearance a penetrating component.

The line with $b=1.9$ corresponds to the scaling in the spectra. This value  is obtained by averaging the experimental values  'b'  in the range $N_e=10^5$--$10^6$. The scaling behavior of the $\gamma$-rays spectra in this energy region is confirmed by LHC data in pp-interactions at the beam energies 1--7 TeV \cite{lhc1,lhc2}. It can be assumed that the scaling in pp is preserved at higher energies \cite{lhc3}. Accounting for nuclear effects in the quark-gluon string model \cite{dun_mq1,dun_mq2,dun_mq3,dun_mq4,dun_mq5} also does not violate scaling in the range of 1--100 PeV.  These model calculations, performed specifically for the conditions of the {\it HADRON} experiment, show that the slope of the spectra should not change in the entire energy range and has the same value $b=1.9$. Therefore, the horizontal line in figure \ref{sp_b_r}  can be considered as a theoretical prediction.

The statistics of events for $\gamma$-families with $\Sigma{E_{\gamma}}\geq10$ TeV, the number of $\gamma$-rays in them and slopes of $E_{\gamma}$-spectra are presented in table \ref{t:1}.

\begin{table}[t]  
\caption{ Statistics of families\texorpdfstring{$N_f$}{Lg} with \texorpdfstring{$\sum E_{\gamma}\geq10$}{Lg} TeV,  families and
\texorpdfstring{$\gamma$}{Lg}-rays with \texorpdfstring{$E_{\gamma}\geq6$}{Lg} TeV and slopes \texorpdfstring{$\bar{b}$}{Lg}  of the  \texorpdfstring{$E_{\gamma}$}{Lg}-spectra for different intervals \texorpdfstring{$\lg(N_e)$}{Lg}.}
\label{t:1}
\begin{tabular}{c c c c c c c }  
\hline 
$\lg(N_e)$ & $N_f$ & $N_f (E_{\gamma}\geq 6 \text{TeV})$  & 
$n_{\gamma}(E_{\gamma}\geq 6  \text{TeV}) $  & $ N_f^{\text{halo}}$ & $<b>^{\text{halo}} $  & $<b>$\\ 
 \hline
 4.93--5.18 & 134 & 119 & 175 & 0 & - & $1.88\pm0.12$\\
 5.18--5.43 & 182 & 164 & 248 & 0 &-& $1.96\pm0.12$\\
 5.43--5.68 & 198 & 181 & 300 & 2 &-& $1.70\pm0.08$\\
 5.68--5.93 & 192 & 176 & 314 & 1 &-& $1.75\pm0.08$\\
 5.93--6.18 & 175 & 154 & 290 & 0 & - & $2.03\pm0.11$\\
 6.18--6.43 & 168 & 149 & 393 & 3 & - & $1.66\pm0.07$\\
 6.43--6.68 &  95 & 91 & 315 & 10 & $1.03\pm0.07$ & $1.43\pm0.06$\\
 6.68--6.93 &  47 & 46 & 194    & 4 & - & $1.40\pm0.08$\\
 6.93--7.18 &  17 & 16 & 62 & 2 & - & $1.13\pm0.12$\\
 7.18--7.43 &  11 & 11 & 89 & 2 & - & $1.48\pm0.15$\\
 7.43--7.68 &   0 &  0 &  0 & 0 & - & -\\
 7.68--7.93&   4  &  4 & 56 & 1 & - & $2.10\pm0.33$\\ 
\hline
\end{tabular}
\end{table}

It is important to note that in part of these events there is  a diffuse spots of darkening  in the center of a $\gamma-$families which named halo \cite{pyatov}.
The statistic of the $\gamma$-families with halo includes only 25 events because of them  high energy $\sum E_{\gamma}\geq250$ TeV. Them spectrum   has a maximum within the range $N_e=2.7$--$4.8\cdot10^6$. The spectrum slope for the  $\gamma$-families with halo exhibits the minimum value $b=1.03\pm0.07$. In addition, the multiplicity of these families are almost four times more than in other events: $<n_{\gamma}^{halo}>=24.8\pm4.1$ and $<n_{\gamma}>=6.6\pm2.1$ for families without halo.
In the model calculations the appearance of halo in the $\gamma$-families explain  with the EAS created mainly by primary protons \cite{pyatov}.

The obtained data quite reliably exclude the option scaling of hadron spectra in the region of the EAS knee.  In the range $N_e=10^6$--$10^8$ there take place a systematic dependence of the  slope of the spectra, which is approximated by a polynomial of the third degree $Y=-72.61412+38.67686 X-5.94488X^2+0.31637X^3$ ($X=b, Y=\log(N_e)$). This approximation represented by a solid curve in figure \ref{sp_b_r}. As it follows from the data presented in the table \ref{t:1} in the region $N_e\sim10^7$ the difference between the theoretical and experimental slopes of the spectra is $\sim6\sigma$.

\begin{figure}[t]
\begin{center}
\begin{minipage}[t]{0.45\linewidth}
\includegraphics[width=8cm,height=6cm]{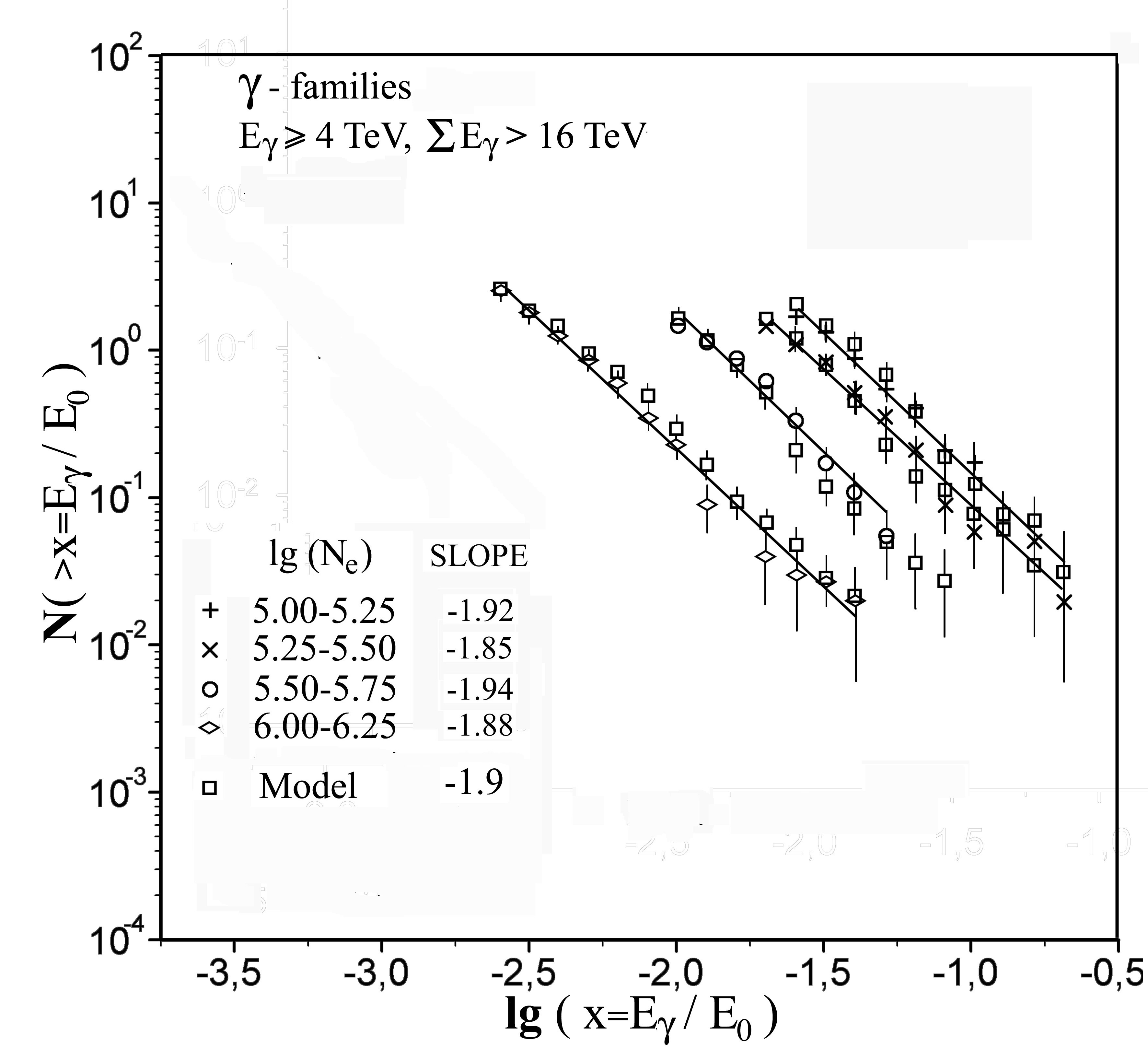}
\caption{Spectra of \texorpdfstring{$\gamma$}{Lg}-rays before the first break of the CR spectrum in comparison with model spectra. }
\label{f:7_l}
\end{minipage}
\hfill
\begin{minipage}[t]{0.45\linewidth}
\includegraphics[width=7cm,height=6cm]{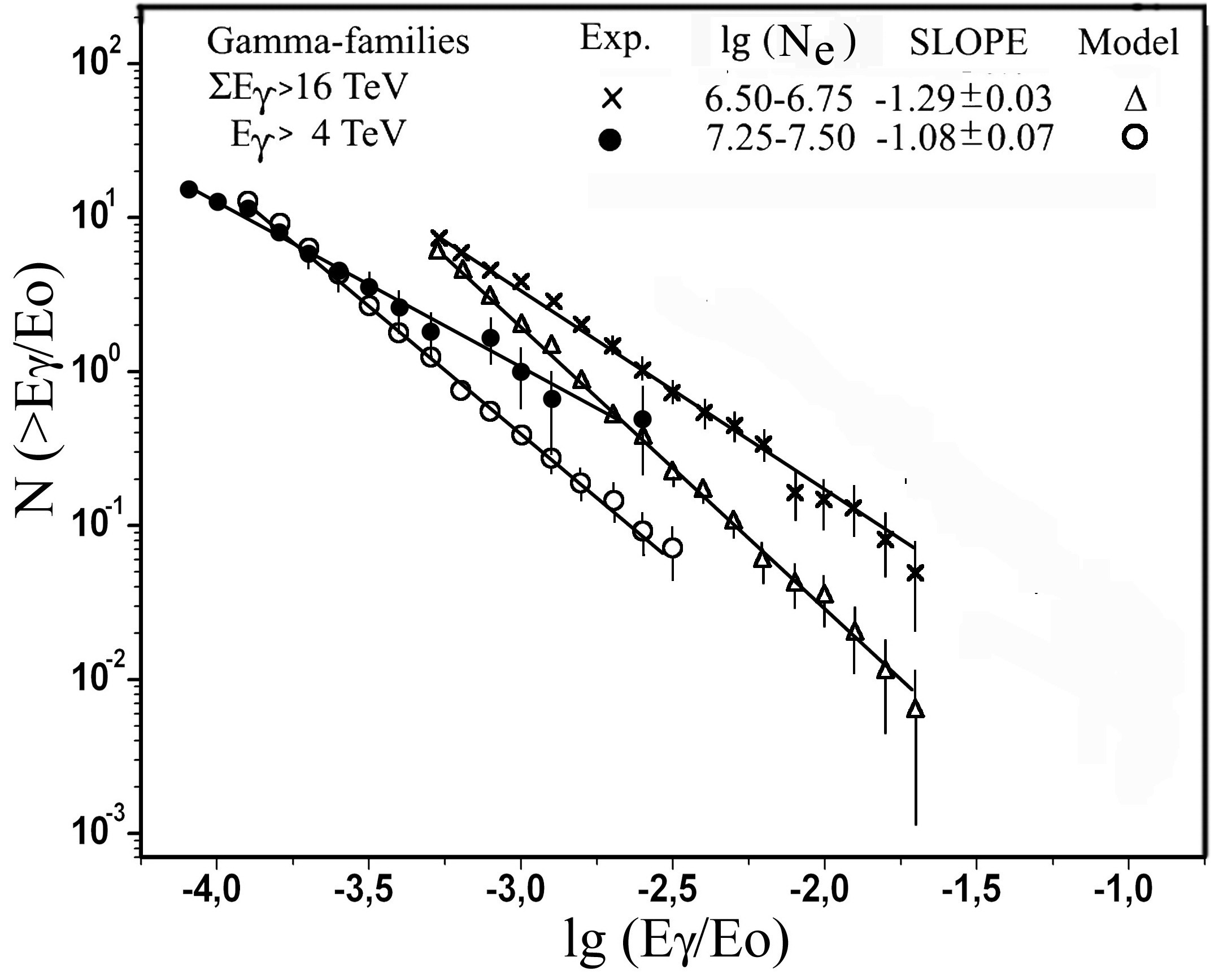}
\caption{The comparison of the experimental and model spectra of \texorpdfstring{$\gamma$}{Lg}-rays in the region of the scaling   violation. }
\label{f:7_r}
\end{minipage}
\end{center}
\end{figure}

At figure \ref{f:7_l} and \ref{f:7_r} experimental and model spectra are compared for a slightly different selection of events by the criteria 
$\Sigma{E_{\gamma}}\geq16$ TeV, $E_{\gamma}\geq4$ TeV. The figure \ref{f:7_l} shows the spectra in the area before the scaling violation. The experimental and model spectra fairly good correspond each other both on slopes and multiplicities. Solid straight lines correspond to the approximation of the experimental data, their slopes are shown in the figure. The figure \ref{f:7_r} shows two  spectra from the region of the scaling  violation. In this case the slopes of the  model and experimental spectra differ significantly.

To verify scaling violation, tests were performed  the absence of errors in the energies of $\gamma$-quanta and the slopes of the spectra.

\begin{figure}[t]
\centering\includegraphics[width=7cm]{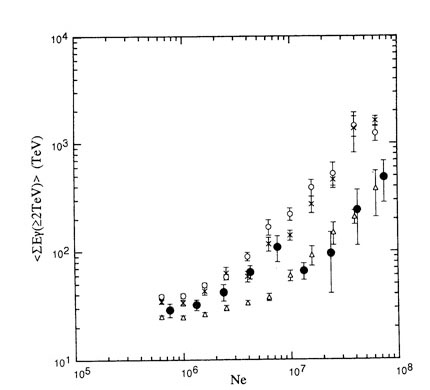}
\caption{\texorpdfstring{$N_e$}{Lg} dependence of the  total energy  for \texorpdfstring{$\gamma$}{Lg}-families. \texorpdfstring{$\bullet$}{Lg} --- experimental data, Model predictions: \texorpdfstring{$\circ$}{Lg} --- light mass composition (A), \texorpdfstring{$\ast$}{Lg} --- heavy mass composition (A), 
\texorpdfstring{$\bigtriangleup$}{Lg} --- Centauro B model.}
\label{smeg}
\end{figure}

First of all, the conditions for the development of x-ray films  were checked using D-labels. The darkening of the film background and labels were compared for all the films. The average value of label darkening  equals to $D_{m}=0.42\pm0.07$, which corresponds to normal conditions of film processing.No films with a substantial deviation from this value were found.

Search for errors  in the energy determination also gave the negative result. It follows from the  calibration curve in figure \ref{dolg} that values of $E_{\gamma}$ for the most energetic cascades were not distorted.

Along with the violation of scaling, it is interesting to note that it's region $N_e=10^6$--$10^8$ coincides with the area of restructuring of the EAS spectrum.  Given the importance of this conclusion, additional methodical checks of this result were performed.

Figure \ref{smeg} demonstrates the $N_e$ dependence of the total energy $\Sigma E_{\gamma}$ for $\gamma$-families, which was obtained for more high statistic.

The experimental data were compared with calculation results for three models \cite{tamada_sum}. In figure \ref{smeg}, data calculated  by two versions of the model A based on accelerator experiments are presented for light and heavy CR mass composition. Model B (Centauro model) is exotic. The calculations according to all these models testifies to a monotonous dependence $\Sigma{E_{\gamma}}$ on $N_e$. No model describes experimental data in the entire range of $N_e$.

The total energy of $\Sigma{E_{\gamma}}$  rise within the interval $N_e =10^6$--$10^7$ both for experiment and  for models. There are two reasons for the growth of experimental values: an increase in primary energy and a violation of scaling. In theoretical models, there is no scaling disturbance and all growth is associated with an increase in primary energy. For models A and B, the energy transferred to the secondary $\gamma$-quanta differs by almost an order of magnitude.

Twofold energy reduction from 100 to 50 TeV in the region above $N_e=10^7$  confirms the local nature  the penetrating component.

\section{Discussion.} \label{disc}

The interpretation of the knee is remains ambiguous up to now.   It is sufficiently to say that the fraction of primary protons in the knee region, which was obtained in different studies, may distinguished in a few times. As result   the break at the energy $3\cdot 10^{15}$ eV  associates with break in either the lightest CR  component (protons)  \cite{grande} or  with the   heaviest nuclei  \cite{tibet}. The reasons in difference of interpretations lies in the different methods and models used for receiving these data.   Herewith it should be noted that the nuclear component is mainly concentrated in the EAS cores, so the data of {\it TIBET} experiment seem more informative. The authors \cite{tibet} conclude that the break at $N_e=10^6$ is formed either by a heavy component from a close single source \cite{erl_w1}, or under certain acceleration conditions from the radiation of many galactic sources. These considerations are related to the results we have obtained.

In contrast to previous experiments with lead chambers, here we analyze the penetrating component formed in the atmosphere. The families $\gamma$-quanta ($\pi^0$) are formed in the atmosphere at heights of 3--4 kilometers above the installation.  In this case, the effect  birth of unstable heavy particles on the characteristics of the EAS is minimal. In particular the role of charmed particles in this case is negligibly small  and cannot explain the observed effects. Them lifetimes  are $10^{-12}$--$10^{-13}$ s and therefore the decay length amount tens of meters. In lead chambers, having the size of meters, the increased cross-section of the birth of enchanted particles can lead to the tightening of the cascade. In the atmosphere the charmed particles must mainly decay before reaching the level of observation. In addition, the birth cross section of the  charmed particles on the  air nuclei should be less than on lead. At last, the cross section grows monotonously, but the break supposes the threshold nature of the effect.

The special calculations which were made for {\it HADRON} experiment conditions  confirm this conclusion \cite{dun_abs}. It was shown that increasing the length of the absorption of hadrons in lead XREC \cite{pam_abs} can only be explained by the increase in the cross-section of the birth of charm in the lead up to $50\pm10\%$ of the full inelastic cross-section at an energy of about 75 TeV. In addition, it is shown that the introduction of charmed particles in the atmosphere cascade  has no effect on the characteristics of EAS at least up to $N_e\sim10^8$.
At the same time, the $N_e$ dependence of the parameter s in figure \ref{ldf_r} indicates a significant  steepening of the showers LDF around $N_e=10^7$. It means that  nevertheless the characteristics of EAS change exactly in the area of energy we are interested in and this change is not related to the leading charm.

Thus, the data of the {\it HADRON} experiment allow us to conclude that the particles of the penetrating component are not products of secondary interactions in the atmosphere, but are present in the primary radiation. The importance of this conclusion lies in the necessary condition for the stability of such particles. There is only one option for known strongly interacting particles --- it is nuclei, among which the most penetrating are protons.

Another important feature of the CR spectrum is the sharpness of its break at the energy 3 PeV. We have established in the previous section that the change of interaction characteristics can't explain this effect. Remains to search astrophysical option.

Astrophysical models  predict  the  CR spectrum fall  for two reasons. As a result of the  diffusion  CR partly can  leaving the Galaxy and with energy increase  the efficiency of their acceleration may to decrease . However, both diffusion and energy models suggest a smooth increase in the slope of the CR spectrum.

A sharp break can occur due to a sharp cut of the nuclear components of the Galactic origin. The traditional explanation of the break at $N_e=10^6$ by break in proton spectrum is not confirmed by the experimental data because it  not consistent with the grows rigidity  the hadron spectra in the range of $N_e=10^6$--$10^7$.  If we assume that for $N_e=10^6$ we are dealing with a break in the Fe nuclei spectrum, then the scaling violation at $N_e>10^6$  has to be associated with a new proton component, and the softening of the spectra in the region $N_e=10^7$--$10^8$ associated with subsequent weighting of the nuclear components.

\section{ Conclusions.}

The energy spectra of $\gamma$-quanta ($\pi^0$) in the knee region $E_0=10^{15}$--$10^{17}$ eV were obtained for the first time using an x-ray emulsion chamber as a detector of EAS cores in hybrid {\it HADRON} experiment.

In the bounded region of primary energies 3--100 PeV ($N_e=10^6$--$10^8$)  a scaling violation of the hadron spectra is happened. The slope of the spectra 'b' ($N(>E_{\gamma})\sim E_{\gamma}^{-b}$) is reduced in the region of $N_e=10^6$--$10^7$ and then in the range of $N_e=10^7$--$10^8$ is returned to the original values. The high values of the hadron energies in the region $N_e=10^6$--$10^8$ means the appearance here  the penetrating EAS component.

The formal reason for the scaling violation of the hadron spectra is the increase in the proportion of $\gamma$-families with extremely rigid spectra (slope $<b>=1.05\pm0.06$ insted $b\sim1.9$). The share of such events reaches $\sim30\%$ in the range $N_e=3\cdot10^6$--$10^7$. It is within this group of events that $\gamma$-families with halo appear. Monte Carlo calculations show that these events appear in the EAS generated by protons \cite{pyatov}.

The most important conclusion is that the penetrating component is not born in the lead chamber but is presented in the primary cosmic rays. The consequence of this is the requirement stability the particles forming a penetrating component, which sharply narrows the list of possible candidates. There is only one such variant among the known  strongly interacting particles --- nuclei, among which protons are the most penetrating component. So we should assume the appearance of a new CR component which consist mainly of protons for the energies about $\sim10$ PeV and heavier nuclei in the energy range of 10--100 PeV.

Such a model clearly requires astrophysical justification. The main doubt is related to the incomprehensible correlation between the intensity and energy of the galactic and new CR component. If the new component is of extragalactic origin, the correlation of these parameters looks extremely unconvincing. The model of a single close source \cite{erl_w1} is more preferable but also raises doubts mainly because of the need large proportion of protons in the nuclear composition of the radiation.

It seems that the data presented here make look for more convincing ways to interpret the knee in the EAS spectrum.

\section{ Acknowledgments.}

The  {\it HADRON} experiment was fulfills by efforts of a large team of participants from the LPI and the high mountain Tien-Shan station of LPI. The author is sincerely grateful to them for the many years fruitful collaboration.

The author  is indebted also to the staff of the  \textit{PAMIR} experiment with whom he cooperated for many years. Their  support he permanently felt during his work at Pamir and Tien Shan.

The  author expresses his particular gratitude Gabriel I. Merzon, who made a number of important comments and gave the invaluable help in  preparing  the manuscript.

This study was supported by  Russian Foundation for Basic Research (project no. N98-02-16942) and by the programs  "Neutrino physics and neutrino astrophysics" and "The fundamental properties of matter and astrophysics" of the Presidium of the Russian Academy of Sciences.

\bibliography{aipsamp}

\begin{thebibliography}{99}\parindent=8truemm
\itemsep -1mm

\bibitem{dremin} I.M.DreminV.I., V.I.Yakovlev,{\it Charm in cosmic rays (The long-flying component of EAS cores)}, Astroparticle Physics, Volume 26, Issue 1, August 2006, Pages 1--9

\bibitem{pam_abs} Amineva T.P., Ivanenko I.P., Il'ina N.P. et al.,{\it Some features of the absorption of high-energy hadrons in 110-cm lead XREC}, Izv. AN SSSR, 1989, ser. phys., {\bf v.53} N2, p. 277

\bibitem{christ}   G.V.Kulikov, G.B.Khristiansen,{\it On the size spectrum of extesive air showers}, JETP, {\bf 35, 3(9)}(1958) p. 635

\bibitem{icetop} M.G.Aartsen, R.Abbasi, Y.Abdou et al. (IceCube collaboration),{\it Measurement of the cosmic ray energy spectrum with IceTop-73}, PHYSICAL REVIEW D 88(4), 042004(15) (2013)

\bibitem{trudi109} V.S.Aseykin, V.P.Bobova et. al, Cosmic rays and high-energy nuclear interactions, Trudy FIAN, 1979, v. 109, p. 3--185

\bibitem{trudi154} L. G. Afanasjeva, S. A. Azimov, Z. A. Azimov et al.(PAMIR Collaboration), {\it Investigation of the nuclear interactions in the energy range $10^{14}$--$10^{17}$ eV by x-ray emulsion chambers in cosmic rays}, Trudy FIAN, 1984, v. 154, p. 3--141

\bibitem{rapoport} Rappoport I.D.,{\it Photographic method for detecting dense showers of charged particles}, J. Exptl. Theoret. Phys. (U.S.S.R.), {\bf 34(7)}, p.998--1000 (1958)

\bibitem{ivanenko} Belyaev A.A., Ivanenko I.P. et al.,{\it "Electron-photon cascades in cosmic rays at ultrahigh energies."}, Moscow, Nauka, 1980, p. 3--308

\bibitem{NKG} Kamata K., Nishimura J.,{\it The lateral and the angular structure functions of electron showers}, Progr. Theor. Phys. 1958. V.6, P.93

\bibitem{lhc1} O. Adriani et al.,{\it Production spectra of zero-degree neutral particles measured by the LHCf experiment}, Phys. Rev. D 94, 032007 (2016)

\bibitem{lhc2} T. Sako, for the LHCf Collaboration,{\it LHCf Measurements of Very Forward Particles at LHC}, XVI International Symposium on Very High Energy Cosmic Ray Interactions ISVHECRI 2010, Batavia, IL, USA (28 June 2 July 2010); arXiv:1010.0195 [hep-ex] (or arXiv:1010.0195v1 [hep-ex])

\bibitem{lhc3} Christian Klein-Boesing, Larry McLerran,{\it Geometrical Scaling of Direct-Photon Production in Hadron Collisions from RHIC to the LHC}, Phys. Lett. B734 282--285 (2014); DOI:	10.1016/j.physletb.2014.05.063; arXiv:1403.1174 [nucl-th]

\bibitem{dun_mq1} Dunaevsky A.M., Pashkov S.V., Slavatinskiy S.A.,{\it Properties of hadron nuclear interactions at $10^{15}$--$10^{17}$ eV}, Proc. of 3 International Symposium on Very High Energy Cosmic Ray Interactions (ISVHECRI), Tokyo, 1984, P.178--198.

\bibitem{dun_mq2} A.M.Dunaevsky, S.V.Pashkov, R.A.Muhamedshin et al.,{\it Calculations of high-energy nuclear-electromagnetic cascades.}, Trudy FIAN, 1984, v. 154, p. 142--217

\bibitem{dun_mq3} Dunaevsky A.M., PlutaM., Slavatinsky S.A.,{\it Transverse momentum, scaling violation in hN interaction and composition of primary cosmic rays at $10^{15}$ eV}, Proc. of 5 International Symposium on Very High Energy Cosmic Ray Interactions (ISVHECRI), Lodz, 1988, P. 143--160.

\bibitem{dun_mq4} Dunaevsky A.D., Krutikova N.P., Slavatinsky S.A.,{\it Simulation of gamma-families accompanied by EAS},  22 ICRC, Dublin, {\bf 4}, p.133--136, 1991

\bibitem{dun_mq5} J.N. Capdevielle, A.M. Dunaevsky, S.A. Karpova, N.P. Krutikova and S.A. Slavatinsky,{\it Total inelasticity in pN interactions at $10^3$--$10^4$ TeV}, J.Phys. G: Nucl. Part. Phys., 20 (1994) 947--959.

\bibitem{pyatov} R.A. Mukhamedshin, V.S. Puchkov, S.E. Pyatovsky, S.B. Shaulov,{\it Analysis of gamma-ray families with halos and estimation of mass composition of primary cosmic radiation at energies 100 PeV}, Astroparticle Physics 102 (2018) 32-38

\bibitem{tamada_sum} M.Tamada, {\it	Air-shower-triggered families: simulation calculation and its comparison with experimental data}, J. Phys. G: Nucl. Part. Phys. {\bf 20} (1994) 487--505

\bibitem{grande} W.D. Apel, J.C. Arteaga-Velazquez, K. Bekk et al.,{\it KASCADE-Grande measurements of energy spectra for elemental groups of cosmic rays}, Astroparticle Physics 47 (2013) p.54-66

\bibitem{tibet} The Tibet AS Collaboration: M. Amenomori, X. J. Bi, D. Chen et al.,{\it Cosmic-ray energy spectrum around the knee observed with the Tibet air-shower experiment},  Astrophys. Space Sci. Trans., 7, p.15-20, 2011 www.astrophys-space-sci-trans.net/7/15/2011/  Author(s) 2011. This work is distributed under the Creative Commons Attribution 3.0 License.

\bibitem{dun_abs} A.M.Dunaevsky, L.G.Sveshnikova, N.P.Krutikova, S.G.Karpova,{\it Simulation of nuclear-electromagnetic cascades initiated by superhigh-energy hadrons in lead and the earth's atmosphere taking into account the generation of charmed particles}, Preprint FIAN 1995, N 18, p. 3--29

\bibitem{erl_w1} A.D.Erlykin, A.W.Wolfendale,{\it A single source of cosmic rays in the range $10^{15}$ -- $10^{16}$ eV}, J. Phys. G:Nucl. Part. Phys. {\bf 23}(1997)979--989
\end{thebibliography}

\end{document}